\documentclass[prl,aps,superscriptaddress, showpacs, twocolumn]{revtex4}

\usepackage{amsmath}
\usepackage{amssymb}
\usepackage[dvips]{graphicx}
\usepackage{type1cm}

\begin{document}

\title{Operational determination of multi-qubit entanglement classes via tuning of local operations}

\date{\today}

\author{T. Bastin}
\affiliation{Institut de Physique Nucl\'eaire, Atomique et de
Spectroscopie, Universit\'e de Li\`ege, 4000 Li\`ege, Belgium}

\author{C. Thiel}
\affiliation{Institut f\"ur Optik, Information und Photonik,
Max-Planck Forschungsgruppe, Universit\"at Erlangen-N\"urnberg,
91058 Erlangen, Germany}

\author{J. von Zanthier}
\affiliation{Institut f\"ur Optik, Information und Photonik,
Max-Planck Forschungsgruppe, Universit\"at Erlangen-N\"urnberg,
91058 Erlangen, Germany}

\author{L. Lamata}
\affiliation{Max-Planck-Institut f\"ur Quantenoptik, Hans-Kopfermann-Strasse 1, 85748 Garching, Germany}
\affiliation{Instituto de Matem\'aticas y F\'{\i}sica Fundamental,
CSIC, Serrano 113-bis, 28006 Madrid, Spain}

\author{E. Solano}
\affiliation{Departamento de Qu\'{\i}mica F\'{\i}sica, Universidad del Pa\'{\i}s Vasco - Euskal Herriko Unibertsitatea, Apdo.\ 644, 48080 Bilbao, Spain}

\author{G. S. Agarwal}
\affiliation{Department of Physics, Oklahoma State University,
Stillwater, OK 74078-3072, USA}

\begin{abstract}
We present a physical setup with which it
is possible to produce arbitrary symmetric long-lived multiqubit entangled states in the internal
ground levels of photon emitters, including the paradigmatic GHZ and W states. In the case of three
emitters, where each tripartite entangled state belongs to one of two well-defined entanglement
classes, we prove a one-to-one correspondence between well-defined sets of experimental parameters, i.e., locally tunable polarizer
orientations, and multiqubit entanglement classes inside the symmetric subspace.
\end{abstract}


\maketitle

Entanglement is a distinctive property of quantum physics associated
with the nonseparable character of multipartite quantum
systems. For the case of two-qubit systems,
entanglement is well understood and can be precisely
quantified~\cite{Woo98}.
Apart from the trivial disentangled case,
three qubits possess two genuine tripartite inequivalent
entanglement classes~\cite{Dur00, Aci01}.
Efforts have been done recently towards higher number of qubits~\cite{Ver02,Lam07,Che06},
including an inductive method~\cite{Lam06}, though so far no comprehensive and scalable classification has been developed.
In this letter, we introduce a physical setup consisting of $N$
emitters, incoherently radiating single photons that may be absorbed
remotely by detectors equipped with polarizers and producing
long-lived multiqubit entangled states among the emitters. We show that it is
possible to associate well-defined sets of locally tuned polarizer
orientations with multiqubit entanglement classes, allowing their
monitoring in an operational manner. Hereby, multipath quantum
interferences, associated with qubit permutation symmetry, play a
key role in explaining the underlying physics.

\begin{figure}[t]
\begin{center}
\vspace*{0.8cm} \noindent\mbox{\includegraphics[width=6.8cm, bb=170 335 680 760, clip=true]{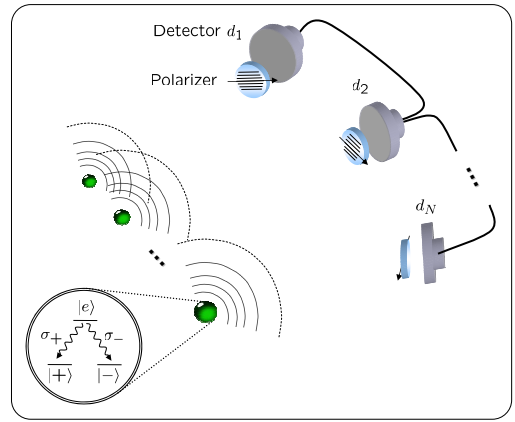}}
\end{center}
\vspace{0cm} \caption{(color online). Proposed experimental arrangement. $N$
excited emitters are aligned in a row, each of them defining a
three-level $\Lambda$-system. A long-lived entangled state is
obtained in the $N$ emitter qubits after detecting the $N$
spontaneously emitted photons with $N$ detectors equipped with
polarizers. The final $N$-qubit state is tuned and determined by the
polarizer orientations.} \label{ionArrangment}
\end{figure}

\begin{figure}
\begin{center}
\mbox{\includegraphics[width=8.5cm, bb=65 330 775 790,
clip=true]{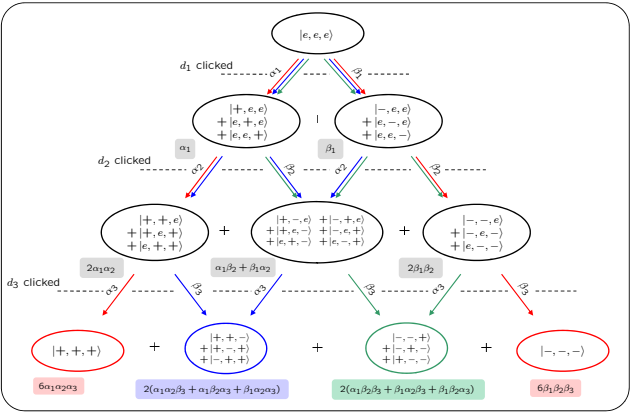}}
\end{center}
\vspace{-0.4cm} \caption{(color online). Pyramid of entanglement paths for the case
of 3 emitters initialized in the excited state $|e,e,e\rangle$. The figure illustrates
the intermediate states, horizontally displayed, during three
successive photodetection events realized by three detectors $d_1$,
$d_2$, and $d_3$, equipped with general elliptical polarizers oriented along $\boldsymbol{\epsilon}_1$,
$\boldsymbol{\epsilon}_2$ and $\boldsymbol{\epsilon}_3$ ($\boldsymbol{\epsilon}_i \equiv
\alpha_i\boldsymbol{\sigma}_+ + \beta_i\boldsymbol{\sigma}_-$), respectively. After each detection,
the unnormalized global state of the system is the sum of all states
in the circles weighed by the tagged prefactors. Left-down arrows
denote $\boldsymbol{\sigma}_+$ transitions, and right-down arrows represent
$\boldsymbol{\sigma}_-$ ones. The final state components are shown
inside colored frames and the colored arrows represent the quantum
paths leading to these different components. Only the red circle
states are obtained via a single quantum path, while the blue and
green ones are the result of three different interfering quantum
paths.} \label{stateEvolution}
\end{figure}

We consider a chain of $N$ equally separated single photon emitters,
say trapped neutral atoms, trapped ions, quantum dots, or any other
equivalent physical system with access to similar behaviour. Each
emitter defines a three-level $\Lambda$ system, where $|e\rangle$
denotes the excited state and the two long-lived sublevels,
$|+\rangle$ and $|-\rangle$, define a qubit. We assume that the
transitions between the excited state and the two lower sublevels
have an equal wavenumber and dipole moment, and that they are
circularly polarized, $\boldsymbol{\sigma}_+$ and $\boldsymbol{\sigma}_-$, respectively.
Figure~1 exemplifies the $N$-emitter case discussed throughout this
paper. All emitters are initially excited and we will study the
cases in which all spontaneously emitted photons are detected by $N$
detectors located in the far-field region, each of them being
equipped with a polarization filter in front. The far-field
detection ensures the erasure of which-way information of the
arriving photons, and the polarizers allow the generation of quantum
superpositions of the lower atomic states when considering arbitrary
polarizations. As a consequence each photodetection event
projects the emitters onto linear combinations of the long-lived
states $|+\rangle$ and $|-\rangle$~\cite{Bli04}. This results at the end in a coherent superposition between the
qubit states $|\pm, \ldots, \pm\rangle$. The indeterminacy of which
detector has projected which emitter in its ground states implies the existence of many
quantum pathways between the initial fully excited atomic state and each of
the final state components. This produces a multipath
quantum interference effect~\cite{Cab99} that we will tune by modifying the
polarizer orientations. We remark that several experimental setups
may be in condition to implement the concepts introduced in
this paper~\cite{Eic93,Mir06,Beu06,Moe07}.

As will be shown explicitly below, it is always possible to find
suitable polarizer orientations to produce any desired state totally symmetric
with respect to permutations of the emitters. Hereby, linear polarizers allow the generation of
a state
of the GHZ type~\cite{Gre90}, the maximally entangled state
\begin{equation}
\label{GHZstate} |{\rm GHZ}_N \rangle \equiv
\frac{1}{\sqrt{2}} \left( |+,\ldots,+\rangle + e^{i \phi}
|-,\ldots,-\rangle \right),
\end{equation}
with arbitrary relative phase $\phi$, or a separable (product)
state%~\cite{Comment1}
\begin{equation}
\label{Sstate} |{\rm S}_N\rangle \equiv
|1_{\phi},\ldots,1_{\phi}\rangle,
\end{equation}
with $|1_{\phi}\rangle \equiv ( |+\rangle + e^{i \phi} |-\rangle ) /
\sqrt{2}$. An intermediate polarizer configuration permits the
generation of the multiqubit state of the W type~\cite{Dur00}, the
weakly entangled state
\begin{equation}
\label{Wstate} |{\rm W}_N\rangle = \frac{1}{\sqrt{N}}
\left( |1_{\phi},0_{\phi},\ldots,0_{\phi}\rangle + \ldots +
|0_{\phi},\ldots,0_{\phi}, 1_{\phi}\rangle \right),
\end{equation}
with $|0_{\phi}\rangle \equiv ( |+\rangle - e^{i \phi} |-\rangle ) /
\sqrt{2}$.
In the 3-qubit case, where each tripartite entangled state belongs to one of two well-defined entanglement classes~\cite{Dur00, Aci01}: GHZ or W
family, it turns out that even the rotation of a single polarizer allows us to
switch from one entanglement class preparation to the other, which are univocally determined by the number of distinct polarizer orientations in the experimental setting. Those polarizer
manipulations can be said to be local with respect to the polarizer/detector
positions, though they are not local with respect to the qubit positions: it is
well known that one cannot convert states from different families into each others
using stochastic local operations and classical communication
(SLOCC)~\cite{ftn1}. Remarkably, the above described local polarization rotations will make possible the transitions
\begin{equation}
\textrm{S class} \leftrightarrow \textrm{W class} \leftrightarrow
\textrm{GHZ class},
\end{equation}
where ``S'' stems from ``separable''.

There are several physical systems where the generation of GHZ and W
states with three or more qubits have been experimentally achieved:
in trapped ions~\cite{Lei05,Haf05}, in Rydberg atoms crossing
microwave cavities~\cite{Rau00}, and in photonic systems~\cite{Pan00, Lan08}.
Furthermore, other multiqubit entangled states have been realized in
different physical setups~\cite{Pan01,Zha04,Sac00,Lei04,Kie07} with
different purposes~\cite{Kok02} and potential applications in
quantum information. Though several paradigmatic entangled states
have been produced in the lab, there is no study, to our knowledge,
that associates operationally given experimental configurations with
multipartite entanglement classes.

The outlined behavior can be understood from the explicit
calculation of the different states of the $N$-emitter system after
the successive photon detection events. A properly located detector
in the far-field with a general elliptical polarizer oriented in the $xy$ plane
of the circularly polarized light along the polarization vector $\boldsymbol{\epsilon} \equiv
\alpha\boldsymbol{\sigma}_+ + \beta\boldsymbol{\sigma}_-$ with arbitrary complex coefficients $\alpha$ and $\beta$ ($|\alpha|^2 + |\beta|^2 = 1$) can implement the
operator action~\cite{Aga02}
\begin{equation}
\label{Dop2} \hat{\cal D}(\boldsymbol{\epsilon}) = \alpha \sum_{j = 1}^N
 |+\rangle_j \langle e| + \beta \sum_{j = 1}^N
|-\rangle_j \langle e| ,
\end{equation}
up to an insignificant prefactor. Here, the sum over $j$ runs
over all emitters and $|\pm\rangle_j\langle e|$ is the
projection operator from state $|e\rangle$ to state $|\pm\rangle$
for emitter $j$. Starting with $N$ emitters in their excited
state $|e,\ldots, e\rangle$, the detection of the $N$ emitted
photons by $N$ detectors with polarizer configuration
$\boldsymbol{\epsilon}_1, \ldots, \boldsymbol{\epsilon}_N$ ($\boldsymbol{\epsilon}_i \equiv
\alpha_i\boldsymbol{\sigma}_+ + \beta_i\boldsymbol{\sigma}_-$) projects the emitter system
onto the final state $|\psi_f\rangle = \hat{\cal D}({\boldsymbol{\epsilon}}_N) \ldots \hat{\cal D}({\boldsymbol{\epsilon}}_1)|e,\ldots, e\rangle$. Due to the symmetry properties of
those operators, the final state as well as all intermediate states
are totally symmetric with respect to permutations of the emitters.
If entanglement is produced at the end of the detection process, we
are ensured that only genuine multipartite entangled states
belonging to the accessible symmetric entanglement classes will be generated. The intermediate states produced in the course of the successive photon
detection events can be ordered in a pyramidal manner, displaying the various
possible quantum paths towards the generation
of the desired final state (see Fig.~\ref{stateEvolution} which exemplifies the case $N=3$). At the end of the $N$ photon detection
process, the system is found in a coherent superposition of
all possible product states of the $N$-qubit system in the
$|\pm\rangle$ basis. All probability amplitudes related to the
different quantum paths add up and yield interference terms (tagged prefactors on Fig.~\ref{stateEvolution}).
The final state $|\psi_f\rangle$ is found
to be a linear combination of all symmetric Dicke states with $k$
$|-\rangle$ excitations~\cite{Comment4,Thi07}, $|D_N(k)\rangle$~:
\begin{equation}
\label{psif}
|\psi_f\rangle = \mathcal{N} \sum_{k=0}^N c_k |D_N(k) \rangle,
\end{equation}
where $\mathcal{N}$ is a normalization prefactor, and
\begin{equation}
\label{ck} c_k = (C_N^k)^{1/2} \sum_{1 \leqslant i_1 \neq \ldots \neq i_N \leqslant N}
    \beta_{i_1} \ldots \beta_{i_k} \alpha_{i_{k+1}} \ldots
    \alpha_{i_N},
\end{equation}
with $C_N^k$ the binomial coefficient of $N$ and $k$. Thereby, we can generate any symmetric state with respect to permutations of the emitters using suitable elliptical polarizer orientations. This is a direct application of Vieta's formulas~\cite{Vieta}~: an arbitrary symmetrical state can be expanded in the Dicke state basis as $\sum_k^N d_k |D_N(k)\rangle$ and is produced in our setup using $K$ polarizers oriented along vectors $\boldsymbol{\epsilon}_i$ with $\alpha_i/\beta_i$ identifying to the $K$ roots of the polynomial $P(z) = \sum_k^N (-1)^{K-k} \sqrt{C_N^k/C_N^K} d_k z^k$ with $K$ the degree of this polynomial, the remaining polarizers oriented along $\boldsymbol{\sigma}_+$.

\begin{figure}
\begin{center}
\noindent\mbox{\includegraphics[width=8.5cm, bb=80
595 770 775, clip=true]{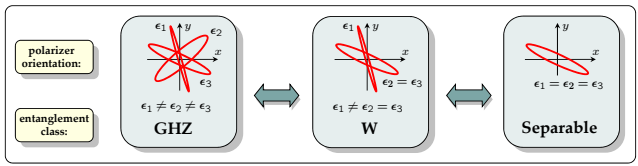}}
\end{center}
\vspace{-0.4cm} \caption{(color online). Operational monitoring of tripartite
entanglement classes by changing the polarization configurations. If
the three polarizer orientations are all different, the final state
belongs to the GHZ class; if two polarizer orientations are different, the
final state belongs to the W class; if all polarizers are
identically oriented, the final state belongs to the S
class.}\label{polarizerFigure}
\end{figure}

For example, the maximally entangled state $|{\rm GHZ}_N\rangle$ of Eq.~(\ref{GHZstate}) is generated with linear polarizers $\boldsymbol{\pi}_{\theta_k} \equiv (e^{-i \theta_k} \boldsymbol{\sigma}_+ + e^{i \theta_k} \boldsymbol{\sigma}_-)/\sqrt{2}$ oriented along angles
\begin{equation}
    \theta_k = \left[ \frac{\pi}{2 N} \right] + \frac{\phi}{2 N} +
    (k - 1) \frac{\pi}{N}, \quad k = 1, \ldots, N ,
\end{equation}
or any configuration resulting from permuting the indices. Here, the term inside the square
brackets is only present for the case of an even number of emitters. The state
$|{\rm GHZ}_N\rangle$ appears naturally when
using all distinct polarizer orientations uniformly shared out over
2$\pi$.

In contrast, we are left with the separable state $|{\rm S}_N\rangle$ of Eq.~(\ref{Sstate}) with linear polarizers all identically oriented
along the angle $\phi/2$~:
\begin{equation}
    \theta_1 = \ldots = \theta_N = \frac{\phi}{2}.
\end{equation}

The state $|{\rm W}_N \rangle$ is also generated with linear polarizers, all
except one oriented identically along $\phi/2$ with the last
orthogonal to the $N-1$ first,
\begin{equation}
    \theta_1 = \ldots = \theta_{N-1} = \frac{\phi}{2}, \quad \theta_N = \frac{\phi}{2} \pm
    \frac{\pi}{2},
\end{equation}
or any configuration obtained with permutation of the indices.

For three emitters, those particular results, associating the states $|{\rm
S}_3\rangle$, $|{\rm W}_3\rangle$, and $|{\rm GHZ}_3\rangle$ with
certain choices of polarization angles, suggest a wider physical
picture. In fact, we can associate in an operational manner specific
polarizer configurations of our proposed physical setup with
the three paradigmatic entanglement classes appearing in the
3-qubit case~: the
number of distinct polarizer orientations identifies univocally the
entanglement class, be S, W, or GHZ, as shown in
Fig.~\ref{polarizerFigure}.

According to D\"ur {\it et
al.}~\cite{Dur00}, the GHZ class is formed by states characterized by a non
vanishing 3-tangle~\cite{Comment2}, the W class by states with
a zero 3-tangle and non-zero single qubit von Neumann entropies,
while the separable state class is characterized by zero values of
these entanglement measures. In our proposed scheme, the 3-tangle $\tau$ of
the final state $|\psi_f\rangle$ reads
\begin{equation}
    \tau = \frac{4}{27} \mathcal{N}^4 |\alpha_1 \beta_2 - \alpha_2 \beta_1|^2 |\alpha_1 \beta_3 - \alpha_3 \beta_1|^2 |\alpha_2 \beta_3 - \alpha_3 \beta_2|^2
\end{equation}
and vanishes only when 2 polarizers are equally
oriented. In this case, the local entropies vanish only when the
third polarizer coincides with the two first.
When switching from a configuration
with three distinct polarizer orientations to a configuration with three
identical ones, via the intermediate case where two of them are
equal, we transit successively from the GHZ class, to the W class,
and end in the S class, as seen
in Fig.~\ref{polarizerFigure}. This remarkably shows the
potentiality of the proposed setup for associating operationally a physical
setting with 3-qubit entanglement SLOCC classes of states symmetric with respect to permutations of the emitters.
These results encourage a possible generalization to the arbitrary $N$-qubit state.

Finally, we give an estimation of the expected fidelity of our
scheme, e.g., for generating the $|{\rm GHZ}_4 \rangle$ state using
4 adjacent ions localized in a linear trap. Assuming a 5~$\mu$m
separation between ions with a 5~nm confinement in the transverse
direction and an azimuthal detection window of 1$^{\circ}$, we
estimate a fidelity of about 90\% for the generation of the 4-qubit
state, largely above the 50\% minimal value for proving genuine
four-qubit entanglement of the state~\cite{See01}. With usual
ellipsometry techniques, polarizer orientation uncertainties can be
made insignificant compared to the finite size effect of the
detection window. With an excitation rate of several tens of
MHz~\cite{Bli04}, the counting rate of the needed fourfold
coincident events can be of the order of several tenths of Hz using
CCD cameras covering a fair area in the detection plane. In general,
the counting rate decreases with the number of qubits. This might
limit the scalability of the scheme as in similar experiments
related to entanglement production~\cite{Pan00, Bli04, Kie07,
Moe07}.

{\it Note added.}--After submission of this work, we became aware of a nice experimental observation of an entire family of four-photon entangled states~\cite{Wie08}.

L.L. thanks Alexander von Humboldt Foundation
for the funding through a Humboldt Research Fellowship. G.S.A.
thanks NSF grant NSF-CCF-0524673 for supporting this collaboration. C.T. and J.v.Z. gratefully acknowledge financial support by the Staedtler foundation.
E.S. thanks to Ikerbasque Foundation, EU EuroSQIP project, and UPV-EHU Grant GIU07/40.
T.B. thanks F. Bastin, P. Mathonet, P. Lecomte and M. Rigo from University of Li\`ege for helpful discussions.

\end{document}